\documentclass[aps,prb,twocolumn,superscriptaddress,showpacs]{revtex4}

\usepackage[dvips]{graphicx}
\usepackage{color}
\usepackage{CJK}
\usepackage{hyperref}
\usepackage{bm}
\renewcommand{\vec}[1]{\boldsymbol{#1}}

\begin{document}

\title{Ground State and Intrinsic Susceptibility of the Kagome Antiferromagnet Vesignieite as seen by $^{51}$V NMR}

\author{J.~A.~Quilliam}
\author{F.~Bert}

\affiliation{Laboratoire de Physique des Solides, Universit\'{e} Paris-Sud 11, UMR CNRS 8502, 91405 Orsay, France}

\author{R.~H.~Colman}
\altaffiliation{Present address: Department of Chemistry, University of Aberdeen, Meston Walk, Aberdeen AB24 3UE, UK}
\author{D.~Boldrin}
\author{A.~S.~Wills}
\affiliation{Department of Chemistry, University College London, 20 Gordon Street, London, WC1H 0AJ, UK}

\author{P.~Mendels}
\affiliation{Laboratoire de Physique des Solides, Universit\'{e} Paris-Sud 11, UMR CNRS 8502, 91405 Orsay, France}

\date{\today}

\begin{abstract}

The intrinsic magnetic susceptibility and local magnetization of the near-kagome quantum magnet vesignieite, Cu$_3$BaV$_2$O$_8$(OH)$_2$, are presented as measured using $^{51}$V NMR.  The NMR line shift gives an accurate measurement of the intrinsic susceptibility of the kagome sites which closely resembles that of the quantum spin liquid herbertsmithite [A.~Olariu \emph{et al.} Phys. Rev. Lett. {\bf 100}, 087202 (2008)].   It is therefore surprising that, at $T_C\simeq 9$ K, a transition to a heterogeneous ground state is observed.  A gradual wipeout of half the NMR intensity indicates a slowly fluctuating spin liquid component and a detailed analysis of the linewidth reveals the onset of static magnetism at the remaining half of the sites.  It is proposed that this transition and unusual ground state originate from a nearby quantum critical point induced by the Dzyaloshinskii-Moriya interaction.

\end{abstract}


\maketitle



Spin-1/2 kagome Heisenberg antiferromagnets (KHAFMs) are generally considered to be the quintessential setting for the discovery of novel quantum spin liquid (QSL) physics.  This has been substantiated by the discovery of a stable spin liquid ground state in the first $S=1/2$ system, herbertsmithite\cite{Mendels2007,Helton2007} with experiments to date indicating gapless excitations.\cite{Helton2007,Olariu2008}   In contrast, a theoretical consensus on the nature of the ground state, even of the ideal case, has been difficult to achieve, with several competing ground states including gapped~\cite{Yan2010} or gapless~\cite{Ran2007} QSLs and valence bond crystals,\cite{Singh2007} supported by different theoretical techniques.  Furthermore, the precise effects of certain non-idealities present in realistic systems, such as in-plane and out-of-plane disorder~\cite{BertdeVries} and the symmetry-allowed Dzyaloshinskii-Moriya (DM) interaction,\cite{Zorko2008} remain important problems to resolve.  Thus, the discovery and characterization of new spin-1/2 KHAFM materials is of fundamental importance to the field of geometrically frustrated quantum magnetism.

The mineral vesignieite, Cu$_3$BaV$_2$O$_8$(OH)$_2$,~\cite{Okamoto2009} exhibits a slightly distorted kagome lattice, with a minute 0.07\% bond-length difference between inequivalent Cu sites~\cite{Colman2011} which both have the same magnetically active orbitals,~\cite{Okamoto2009} making it an excellent approximation of a $S=1/2$ kagome system with nearest neighbor exchange of $J \simeq J' = 53$ K.~\cite{Okamoto2009}  It is therefore a challenging counterpart to the only other such material available, herbertsmithite. 
  A recent in-depth study of new high quality samples of vesignieite has revealed some level of spin freezing at a surprisingly high temperature, $T_C\simeq 9$ K $\simeq J/6$, as seen from a separation between field-cooled (FC) and zero-field-cooled (ZFC) bulk susceptibility, the appearance of weak static magnetism in $\mu$SR experiments and a reduction in the paramagnetic-like contribution to powder neutron diffraction.~\cite{Colman2011}  


\begin{figure}
\begin{center}
\includegraphics[width=3.55in,keepaspectratio=true]{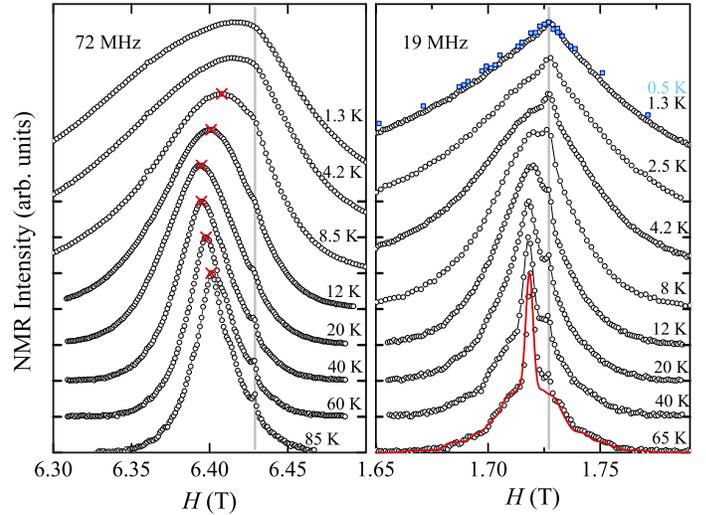}
\caption{(color online) NMR spectra taken at two different frequencies: 19 and 72 MHz.  The spectra are normalized to their maximum value and then evenly spaced to improve visibility (tick marks on the vertical axes indicate the baselines of the spectra).  The spectrum at 0.5 K (blue) is superimposed directly on top of the spectrum at 1.3 K to show that there is very little change in the width below that point.  Gray lines indicate the reference field.  The solid red line is a powder simulation of the spectrum at 65 K (see text).  The maxima of the 72 MHz spectra (red x's) determine the line shift $K(T)$.}
\label{Spectra}
\end{center}
\end{figure}

A crucial next step in our understanding of KHAFMs such as vesignieite must come from accurate determination of fundamental thermodynamic quantities which can be complicated or even overwhelmed by the presence of defects.~\cite{BertdeVries}  Hence, a local probe such as NMR is particularly valuable with its ability to resolve the intrinsic magnetic susceptibility, $\chi_\mathrm{int}$, of the kagome planes.~\cite{Olariu2008} 
In this report, we apply $^{51}$V NMR measurements to the vesignieite system and obtain a measure of $\chi_\mathrm{int}$ that shows remarkable similarities to the QSL herbertsmithite.  Surprisingly, we also observe weak static magnetic moments, with appreciable quantum fluctuations, below $T_C$.  This frozen component is found to affect 50\% of the hexagons of the kagome lattice, with the remaining half showing slow spin liquid dynamics.


The details of sample preparation have been described in Ref.~\onlinecite{Colman2011}.  The majority of measurements presented here were performed on a sample (sample A) that underwent hydrothermal annealing for 48 hours, the same polycrystalline sample employed previously in $\mu$SR experiments.~\cite{Colman2011}  Further increased annealing times were found to have a negligible effect on sample quality.   A second sample (sample B) that was not hydrothermally annealed does not show a kink at 9 K, or a difference between FC and ZFC susceptibility~\cite{SI} as in Ref.~\onlinecite{Okamoto2009}.  The latter sample was studied with a limited set of measurements to test the effects of disorder.

$^{51}$V NMR measurements were performed by sweeping the field at two distinct frequencies: 19 MHz and 72 MHz.  High $T$ spectra exhibit the typical form of a spin-7/2 quadrupolar nucleus, consisting of a narrow central line on top of a broader feature composed of powder-averaged quadrupolar satellites.  Simulating the spectrum at 65 K and 19 MHz (shown as the red curve in the right panel of Fig.~\ref{Spectra}), results in a quadrupolar splitting, $\nu_Q = 200 \pm 20$ kHz and asymmetry parameter, $\eta \simeq 0.45$, due to the slightly distorted tetrahedral local environment.  Apparent in all spectra is a small, sharp peak, attributed to a $\sim 1$\% non-magnetic parasitic phase.  The evolution in $T$ of the NMR spectra at two different frequencies (magnetic fields) is shown in Fig.~\ref{Spectra}. The induced local magnetic field at the V site may easily be followed down to $T_C$ with the shift, $K$, of the peak value relative to the reference.  Comparison of $K$ with $\chi_\mathrm{macro}$ from 85 K to 300 K, allows us to isolate the intrinsic susceptibility, $\chi_\mathrm{int} = (K-\sigma)/A$, and determine the hyperfine coupling constant $A =  7.7$ kOe/$\mu_B$ and chemical shift of $\sigma = 325$ ppm.  Each $^{51}$V nucleus equally probes six Cu spins on a hexagon, hence the coupling for an individual electron spin is $A/6$.  

\begin{figure}
\begin{center}
\includegraphics[width=3.25in,keepaspectratio=true]{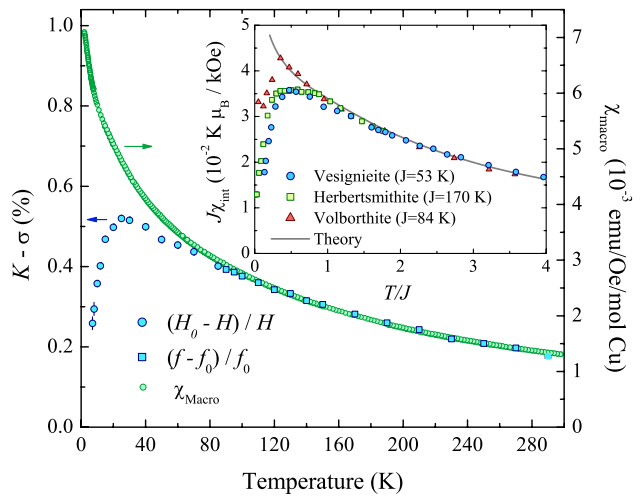}
\caption{(color online) Macroscopic susceptibility taken at 5 T (right axis) and $^{51}$V NMR line shift (left axis) of vesignieite from 9 K to 300 K.  The inset shows a comparison of intrinsic susceptibilities of vesignieite, herbertsmithite~\cite{Olariu2008} and volborthite~\cite{Hiroi2001} with the theoretical kagome susceptibility.~\cite{Misguich2007}  The volborthite shift~\cite{Hiroi2001} was reanalyzed in the same way as the other materials, giving $A = 8.1$ $\mu_B$/kOe and $\sigma = 560$ ppm.}  
\label{KandChiFigure}
\end{center}
\end{figure}

At $T\lesssim 70$ K, $\chi_\mathrm{macro}$ and $\chi_\mathrm{int}$ begin to differ appreciably and by $\sim 25$ K, $\chi_\mathrm{int}$ reaches a maximum value and decreases at lower $T$.  This non-monotonic behavior of the local susceptibility is a recurring feature of magnetic systems on the kagome lattice and is seen in  SrCr$_8$Ga$_4$O$_{19}$ (SCGO),~\cite{Mendels2000} volborthite~\cite{Hiroi2001} and herbertsmithite.~\cite{Olariu2008}   It contrasts sharply with the monotonic increase of $\chi_\mathrm{macro}$, which is often dominated by a defect contribution.~\cite{Olariu2008}  This result substantiates the phenomenological approach of Ref.~\onlinecite{Okamoto2009}, where a Curie tail, $\chi_\mathrm{tail} \propto 1/(T-\theta_\mathrm{W})$, with $\theta_\mathrm{W} = -1.7$ K, equivalent to 7\% isolated $S=1/2$ moments, is subtracted from the total $\chi_\mathrm{macro}$. However, it is unlikely that this contribution can be attributed to simple free spin-1/2 defects in vesignieite, since there is no obvious mechanism available such as the Zn-Cu disorder that occurs in herbertsmithite~\cite{Olariu2008}  hence the origin of this contribution to $\chi_\mathrm{macro}$ remains unclear.

In the inset of Fig.~\ref{KandChiFigure}, we show a comparison of $\chi_\mathrm{int} J$ plotted against $T/J$ for the three available candidate spin-1/2 KHAFMs: vesignieite, symmetric herbertsmithite~\cite{Olariu2008} and the 3\%-distorted system volborthite.~\cite{Hiroi2001}  Foremost, we observe a high degree of similarity between vesignieite (for $T>T_C$) and the QSL, herbertsmithite, with both exhibiting a maximum at $T/J\simeq 0.5$.  This supports the idea that the distortion in vesignieite has little to no impact.  Meanwhile, the least ideal system volborthite~\cite{Hiroi2001} exhibits a more sharply peaked $\chi_\mathrm{int}$ and is surprisingly the best match to theory of the ideal kagome lattice,~\cite{Misguich2007} a correspondence that is likely accidental, resulting from the interplay of the DM interaction and complicated exchange paths.~\cite{Janson2010}  The $\chi_\mathrm{int}(T)$ curves of neither vesignieite nor herbertsmithite match well with theoretical calculations on the ideal KHAFM model~\cite{Misguich2007} which show no downturn until $T/J < 0.2$.  It is tempting to propose the DM interaction, known to be significant in herbertsmithite where $0.044<D/J <0.08$,~\cite{Zorko2008,ElShawish2010} as an explanation for this discrepancy and indeed, exact diagonalization calculations on small clusters show that an out-of-plane $\vec{D}$-vector causes a reduction of $\chi$.~\cite{Rigol2007}  However, for a maximum in $\chi_\mathrm{int}$ to occur at $T/J=0.5$ as seen here, requires an unrealistic $D_\perp > 0.3$.~\cite{Rigol2007}  Thus the effects of the DM interaction on the high-$T$ susceptibility are in fact rather subtle and it is likely that additional model parameters must be introduced to explain $\chi_\mathrm{int}$ of both herbertsmithite and vesignieite.


At a temperature $T_C \simeq 9$ K, two main effects emerge that indicate the presence of a transition to a very unconventional ground state: (1) a  non-paramagnetic increase of the spectral widths (see Fig. 3) and (2) a pronounced loss of NMR intensity (see Fig. 4), culminating in the loss of 50\% of the signal at the lowest $T$.  

\emph{Intensity} -- At $T>T_C$, once corrected for variations in spin-spin relaxation time, $T_2$, the integrated intensity is found to scale well with $1/T$, indicating that we are able to detect 100\% of the nuclei.  Below $T_C$, the intensity begins to drop out gradually, as shown in Fig.~\ref{IntensityFigure}, eventually reaching a constant 50\% of the nuclei.  The missing intensity is lost due to rapid relaxation ($T_1$ or $T_2$) of the nuclear spins as a result of  slowing magnetic fluctuations, as occurs in SCGO~\cite{Mendels2000} and Nd$_3$Ga$_5$SiO$_{14}$~\cite{Zorko2008NGSO} for example.    This intensity does not return at $T\ll T_C$, indicating the presence of persistent slow spin dynamics in the ground state of vesignieite.  Thus, it appears that two inequivalent $V$-sites develop: one senses spin liquid behavior where spins remain dynamic down to low $T$ and the other, always observable, detects spins that are static below $T_C$.  


\emph{Linewidth} -- Unlike the line shift, the linewidth in vesignieite monotonically increases with lower $T$.   In general, the magnetic broadening, $\Delta H$,~\cite{SI} is representative of the distribution of internal magnetic fields, which, in a paramagnetic system, come from induced moments that scale linearly with applied magnetic field, $H_0$.  In the case of a transition to a frozen state with a local static moment ($m$), one should expect a width $\Delta H$ that is largely \emph{independent} of $H_0$.  

\begin{figure}
\begin{center}
\includegraphics[width=3.25in,keepaspectratio=true]{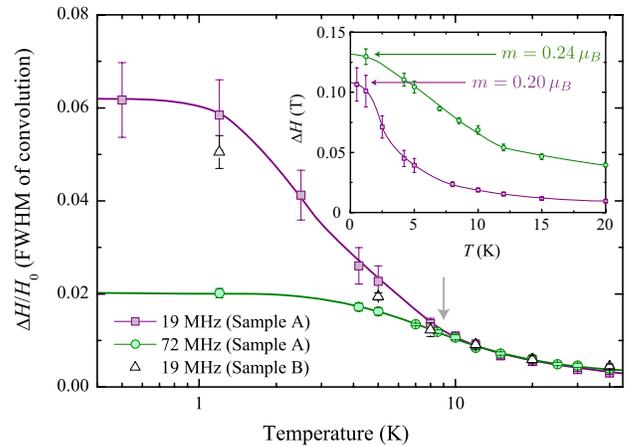}
\caption{(color online) Magnetic broadening, $\Delta H$, for 19 and 72 MHz, scaled by the reference field $H_0$.  Large error bars at low $T$ result from the proximity of the parasitic phase to the central peak.  Sample B, measured at 19 MHz, shows essentially the same result.  The inset shows the unscaled $\Delta H(T)$, which becomes almost independent of $H_0$ at low $T$.  Solid lines are guides to the eye.}
\label{WidthFigure}
\end{center}
\end{figure}

\begin{figure}
\begin{center}
\includegraphics[width=3.25in,keepaspectratio=true]{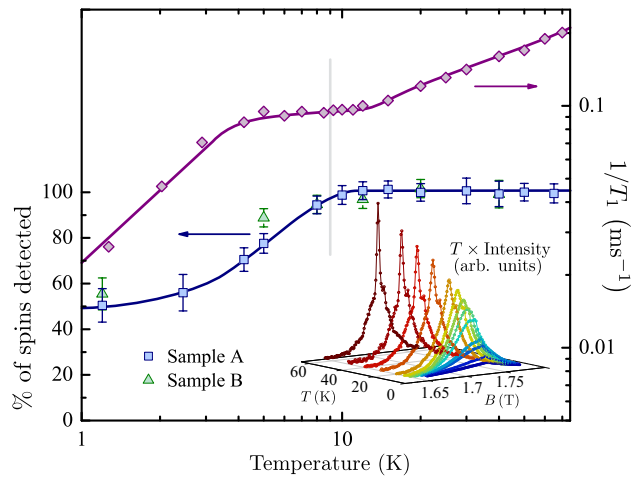}
\caption{(color online) Left axis: percentage of NMR detectable spins.  Right axis: spin-lattice relaxation rate, $T_1^{-1}$, measured at 72 MHz on sample A.  The vertical line indicates the estimated transition temperature. Solid curves are guides to the eye.  The 19 MHz spectra with absolute intensity multiplied by $T$ and corrected for $T_2$ (as much as a 20\% correction at high $T$) are displayed in the inset.  The integrated intensity reflects the number of detected nuclei. }
\label{IntensityFigure}
\end{center}
\end{figure}

In vesignieite, above $T_C$, $\Delta H$ indeed scales linearly with $H_0$.  In this regime, the magnetic broadening is typically the result of a staggered magnetization induced by defects in the kagome lattice.~\cite{IntroFrustMag}  Below $T_C$, $\Delta H$ is no longer found to scale with field as shown in Fig.~\ref{WidthFigure}, indicating an onset of static magnetism.  The development of this frozen moment appears very gradual because, for intermediate $T$ between our base temperature and $T_C$, the spectra contain both static and dynamic components.  Only at the lowest temperatures is the entire spin liquid portion of the sample lost from view, and there $\Delta H$ reaches a largely field-independent limit.

No indication of rectangular spectra, typical of an ordered spin state,~\cite{Bert2005} is seen, suggesting that the spins are frozen randomly or ordered in a  complicated, spin-modulated structure which result in rounded lineshapes.~\cite{Yoshida2011}  $\Delta H$ is mostly saturated below $T=1.3$ K and, under the assumption of randomly frozen spins, represents a static moment, $m = 0.20 \pm 0.02$~$\mu_B$, comparable to the $m\simeq 0.1$~$\mu_B$ estimated from $\mu$SR results in zero field.~\cite{Colman2011}  Even at the lowest $T$, there remains a minor increase in $\Delta H$ and in asymmetry with increasing $H_0$ signaling a residual anisotropic susceptibility of the ordered spins.   At 6.4 T, the moment is then closer to $m\simeq 0.24$ $\mu_B$.    In any case, the static magnetic moment is quite small relative to the total 1 $\mu_B$ Cu$^{2+}$ moment, implying that there are appreciable quantum fluctuations, even in the frozen component of the ground state.

The spin-lattice relaxation rate, $1/T_1$, was obtained with stretched exponential fits to the relaxation curves and a $\beta$ varying smoothly from $\sim 0.55$ to $\sim 0.3$ from high to low $T$.~\cite{SI} Shown in Fig.~\ref{IntensityFigure}, $1/T_1$ exhibits a clear plateau in the vicinity of the transition between $\sim 5$ K and 12 K, representing a broad crossover region.  Below that point, $1/T_1$ becomes steeper in $T$, close to linear, and is then largely probing the ordered state.  Again, no sharp feature is observed at $T_C$, further demonstrating the unusual broad character of the transition and illustrating the combination of dynamic and static spins.  A more typical magnetic transition and critical slowing would give rise to a peak in $T_1^{-1}$.

The results obtained here paint a picture of a transition at $T_C \simeq 9$ K below which a 50\%-50\% mixed static and spin liquid state gradually develops.  This is remarkably similar to the conclusions of $\mu$SR experiments~\cite{Colman2011} that show a 40\% frozen fraction, a minor discrepancy given that the muons and V nuclei are sensitive to different numbers of spins.  In both $\mu$SR~\cite{Colman2011} and NMR studies, the entire sample encounters the same $T_C$, with a slowing down of the spin liquid component occurring in tandem with the appearance of static moments.  This, and the reproducible appearance of two inequivalent sites with equal population, indicate that there is not a macroscopic phase separation in the ground state, but rather a complicated magnetic superstructure occurs.


Despite the fact that vesignieite exhibits a very similar $\chi_\mathrm{int}$ to the structurally perfect KHAFM herbertsmithite, consistent with the minute 0.07\% distortion of its kagome network, it freezes at a $T_C\simeq J/6$, much higher than expected for such a highly frustrated system.  We propose that the similarities in $\chi_\mathrm{int}$ imply that the models of herbertsmithite and vesignieite are quite close, but that an additional DM interaction term places them near to a quantum critical point (QCP), expected to occur at $D_C/J\simeq 0.1$ in the $S=1/2$ case.~\cite{Cepas2008}  While herbertsmithite has $0.044<D/J < 0.08$,~\cite{Zorko2008,ElShawish2010} and thus remains a QSL, vesignieite appears to be just on the other side of the QCP with $D/J \simeq 0.13$, judging by the small static moment of 0.2 $\mu_B$,~\cite{Cepas2008} and $0.1 < D/J < 0.16$ from the width of high-$T$ ESR spectra.~\cite{Zhang2010,ElShawish2010}   This would make our characterization of vesignieite the first experimental support for the theoretical proposal~\cite{Cepas2008} of a quantum critical point sustained by the magnitude of the DM-interaction in kagome physics.


The observed FC-ZFC separation is suggestive of a tendency toward glassiness predicted for a distorted kagome lattice.~\cite{Wang2007}  However, the hysteresis is suppressed by low magnetic fields $\sim 0.5$ T, whereas $T_C \simeq 9$ K is detected with NMR in fields as high as 6.4 T, indicating that vesignieite is not a conventional spin glass.  Comparing samples A and B, we see no change at the local level, suggesting that the transition is not a disorder-driven phenomenon but is intrinsic to the Hamiltonian.  It remains possible that the heterogeneous ground state is a result of pinning of frozen moments by the same disorder that gives rise to the Curie tail in $\chi_\mathrm{Macro}$.  The FC-ZFC splitting is a likely indicator of a weak ferromagnetic out-of-plane canting caused by an in-plane DM interaction, $D_p$,~\cite{Elhajal2002} with low energy barriers that are easily overcome with a small field $\sim 0.1$ T.  Disorder in sample B and perhaps that of Okamoto \emph{et al.}~\cite{Okamoto2009} might effectively break up large ordered domains and further decrease energy barriers, thereby completely erasing the hysteresis.  


The unusual ordering of vesignieite shares some important similarities with two particular quantum magnets.  First, volborthite is also found with $^{51}$V NMR to exhibit a heterogeneous ground state~\cite{Bert2005,Yoshida2011} with two inequivalent $V$ sites, though both show static moments~\cite{Yoshida2011} at variance with our findings.  A similar shallow power law in $T_1^{-1}\propto T$ is seen at low $T$ and is suggestive of a high density of low lying excitations.~\cite{Yoshida2009}  Second, the organic triangular lattice material $\kappa$-(ET)$_2$Cu$_2$(CN)$_3$, which shows QSL physics in zero field, exhibits a field induced transition with a very similar crossover region and power law ($\sim T^{3/2}$) drop in $1/T_1$.~\cite{Shimizu2006}  In that system, the field-induced ground state is also found to be inhomogeneous and only partially ordered with small moments.~\cite{Shimizu2006,Pratt2011}  Thus, we are seeing the emergence of a distinct class of highly frustrated systems that exhibit several common features including heterogeneous ground states, small static moments and low-lying excitations, properties that may be a direct consequence of either DM-induced~\cite{Cepas2008} or field-induced~\cite{Shimizu2006,Pratt2011} quantum criticality.


In conclusion, this local probe study has shed light on the ground state and local susceptibility of the spin-1/2 near-kagome system vesignieite.  While $\chi_\mathrm{int}$ behaves similarly to that of the QSL herbertsmithite, vesignieite is found to partially order as high as $T_C \simeq J/6$.  We have proposed that the DM interaction likely explains this ordering transition and that a proximity to a QCP between spin liquid and N\'{e}el order gives rise to the unusual heterogeneous and dynamic ground state.

\begin{acknowledgments}
We acknowledge informative conversations with O.~C\'{e}pas.  Work was funded by the
ANR-09-JCJC-0093-01 grant.  J. Q. acknowledges support from NSERC.
\end{acknowledgments}

\end{document}